\documentclass[aps,epsf,superscriptaddress,twocolumn,amsmath,amssymb,showpacs,floatfix]{revtex4}

\usepackage{amsmath}
\usepackage{graphicx}
\usepackage{amssymb}

\usepackage{dcolumn}
\usepackage{bm}

\usepackage{upgreek}
\newcommand{\eg}{{\sl e.g.}}
\newcommand{\etal}{{\sl et al.}}

\newcommand{\abt}{{\sl ab initio} thermodynamics}
\newcommand{\magn}{Fe$_3$O$_4$}

\newcommand{\fetw}{Fe$^{\rm 2+}$}
\newcommand{\feth}{Fe$^{\rm 3+}$}
\newcommand{\fea}{Fe$_{\rm A}$}
\newcommand{\feb}{Fe$_{\rm B}$}
\newcommand{\hto}{H$_2$O}
\newcommand{\sqtwo}{$(\sqrt{2}\times \sqrt{2})R45^{\circ}$}

\newcommand{\Os}{O$^{\rm S}$}
\newcommand{\Ow}{O$^{\rm w}$}
\newcommand{\Ooh}{O$^{\rm OH}$}

\makeatletter

\begin{document}
\title{Partial dissociation of water on Fe$_{3}$O$_{4}$(001): adsorbate induced charge and orbital order}

\author{Narasimham Mulakaluri}
\affiliation{Dept. of Earth and Environmental Sciences, University
of Munich, Theresienstr. 41, 80333 Munich, Germany}
\affiliation{Fritz-Haber-Institut der Max-Planck-Gesellschaft,
  Faradayweg 4-6, D-14195 Berlin, Germany}

\author{Rossitza Pentcheva}
\email{pentcheva@lrz.uni-muenchen.de}
\affiliation{Dept. of Earth and Environmental Sciences, University
of Munich, Theresienstr. 41, 80333 Munich, Germany}

\author{Maria Wieland}
\affiliation{Dept. of Earth and Environmental Sciences, University
of Munich, Theresienstr. 41, 80333 Munich, Germany}

\author{Wolfgang Moritz}
\affiliation{Dept. of Earth and Environmental Sciences, University
of Munich, Theresienstr. 41, 80333 Munich, Germany}

\author{Matthias Scheffler}
\affiliation{Fritz-Haber-Institut der Max-Planck-Gesellschaft,
  Faradayweg 4-6, D-14195 Berlin, Germany}

\date{\today}


\begin{abstract}
 The interaction of water with Fe$_3$O$_4$(001) is studied by density functional theory (DFT) calculations including an on-site Coulomb term. For isolated molecules  dissociative adsorption is strongly promoted at surface defect sites, while at higher coverages a
  hydrogen-bonded network forms with alternating molecular and dissociated species.
  This mixed adsorption mode and a suppression of the $(\sqrt{2}\times \sqrt{2})R45^{\circ}$-reconstruction are confirmed by a quantitative low energy electron diffraction (LEED) analysis. Adsorbate induced electron transfer processes  add a new dimension towards understanding the catalytic activity of magnetite(001).
\end{abstract}


\maketitle

Magnetite has attracted continued interest in the past decades due
to its fascinating properties: The high temperature phase is predicted to be a half-metallic
ferrimagnet~\cite{satpathy} with high magnetic ordering temperature.
At around 120~K it undergoes the so called Verwey transition~\cite{verwey}.
Both the type of transition  (metal-to-insulator vs.
semiconductor-to-semiconductor~\cite{park,schrupp}) and the type of
charge (CO) and orbital ordering (OO) at the octahedral iron sites
in the low temperature phase are  subject of an ongoing
debate~\cite{radaelli,leonov,guo,schlappa}.

Besides its applications in magnetic recording and as a
prospective material for spintronics, magnetite
acts as a catalyst \eg\ in environmental redox reactions~\cite{As,Cr},
or in the water gas phase
shift reaction ~\cite{martos}. Typically these reactions take place
in an aqueous environment prompting the need to understand how water
interacts with the Fe$_{3}$O$_{4}$-surface. Water can bind to a
surface in different modes (\eg\ molecular or dissociative) leading to a variety of functional groups that can  affect significantly the surface
properties and availability of reaction sites and result in  a
complex surface chemistry~\cite{henderson02,madey87}.

The catalytic activity of magnetite is typically related to the
presence of both ferrous and ferric iron in its inverse spinel
structure. In the [001]-direction two types of layers alternate:
A-layers with tetrahedral iron (\fea$^{3+}$) and B-layers containing
oxygen and octahedral iron (\feb$^{2+,3+}$).
Both bulk truncations of \magn(001), either with an A- or a
B-layer are polar of type three according to the classification of
Tasker~\cite{tasker}. To explain the origin of the  \sqtwo-reconstruction
previous surface models proposed an ordering of surface defects (\eg~\cite{chambers_001,stanka00}). Recently, DFT
calculations~\cite{pentcheva05,lodziana07} have shown that the
symmetry lowering at \magn(001) is achieved rather by a distorted
B-layer, supported also by surface x-ray
diffraction~\cite{pentcheva05}, LEED~\cite{pentcheva08} and scanning
tunneling microscopy (STM)~\cite{fonin05}.

Despite its importance, only a few studies have addressed the
interaction of water with \magn(001). While initial adsorption was
related to surface defect sites, an extensive hydroxylation  was
reported from x-ray photoemission experiments (XPS) beyond a
threshold pressure of $10^{-3}-10^{-4}$ mbar\cite{Kendelewicz}.
Both this study and temperature programmed desorption
(TPD) measurements\cite{peden} indicate multiple adsorbate sites on the surface.
Molecular dynamics simulations with empirical potentials
~\cite{rustad,kundu} point towards a dissociative adsorption.
However, such studies cannot provide reliable information on the
energetics and underlying electronic phenomena.

Here we address these fundamental questions within a combined DFT
and LEED approach. By varying the coverage and adsorbate
configuration of water molecules we compile a surface phase diagram
in the framework of \abt. The results show that isolated  molecules
dissociate on the clean \magn(001) surface. This process is strongly
favored at oxygen vacancies. With increasing coverage a crossover to
a mixed adsorption (both molecular and dissociative) takes place
that is confirmed in a quantitative LEED analysis. LEED
shows also a strong suppression of the \sqtwo\ surface reconstruction.
Furthermore we find that the adsorbed species invoke electron
transfer processes in the subsurface layers resulting in an unique
charge and orbital order that may  have important
implications on the catalytic activity of the surface.

DFT calculations were performed using
 the full potential linear augmented plane
wave method in the WIEN2K~\cite{wien} implementation. The
generalised gradient approximation (GGA-PBE)~\cite{perdew} of the
exchange-correlation potential is used. Systematic studies~\cite{santraa,santrab} show that the PBE functional gives a good description of hydrogen bonding w.r.t. equilibrium geometries. The error of overbinding which can be as large as
20 meV/bond will not affect our conclusions. Because magnetite is a
strongly correlated material we have explored  the influence of
electronic correlations within LDA/GGA+$U$ ~\cite{anisimov}
using $U=5$~eV and $J=1$~eV~\cite{Uvalue}, similar to  values used for bulk
\magn~\cite{leonov,guo,pinto06}.  The surfaces were modelled in the
supercell geometry with slabs containing seven B-layers and six
A-layers separated by a vacuum region of 10-12 \AA \cite{calcdet}.
The positions of adsorbates and the atoms in the outer two BA-layers were relaxed.

\begin{figure}[t!]
  \centering
  \scalebox{0.4}{\includegraphics{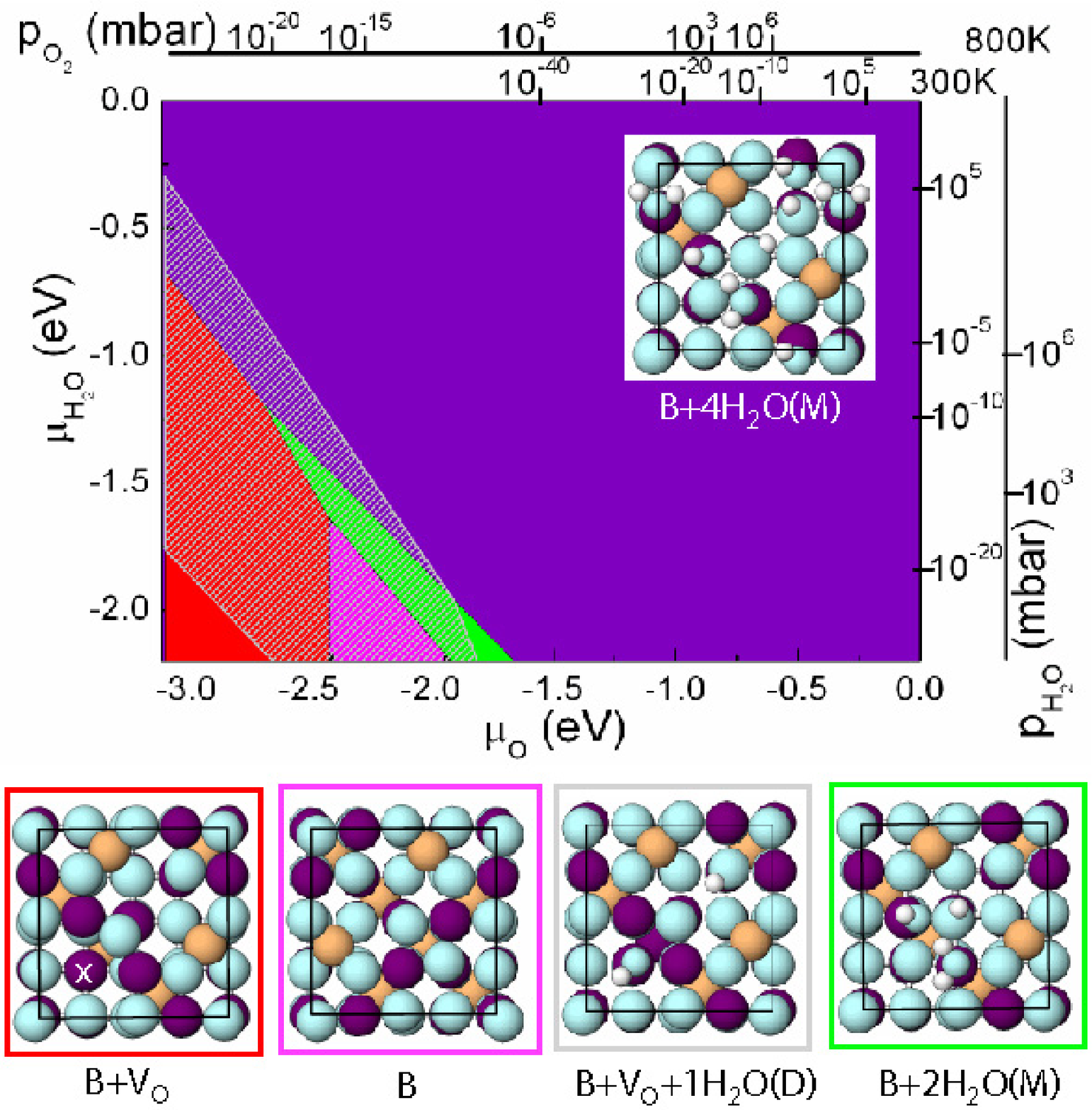}}
  \caption{Bottom view of the surface phase diagram of \magn(001) showing the most stable
  configurations for given $(\mu_{O},\mu_{\rm H_2O})$: B+V$_{\rm O}$ (red), B-layer (magenta), dissociated molecule in an oxygen vacancy B+V$_{\rm O}+1\rm H_2O(D)$ (grey hatched semitransparent area), mixed adsorption of two (green) and four water molecules on a B-layer (purple). The ranges of  $\mu_{O}$ and $\mu_{\rm H_2O}$ correspond to the vapor phase of \hto. $\mu_{O}$ and $\mu_{\rm H_2O}$ have been converted into pressures for $300$ and 800~K.  Additionally, the top views of the most stable configurations are displayed with positions of O, \feb, \fea\ and H marked by cyan, purple, orange and white circles, respectively. In B+V$_{\rm O}$ a white cross marks the position of the vacancy. }
  \label{surf_phase}
\end{figure}

\begin{figure}[t!]
  \centering
  \scalebox{0.42}{\includegraphics{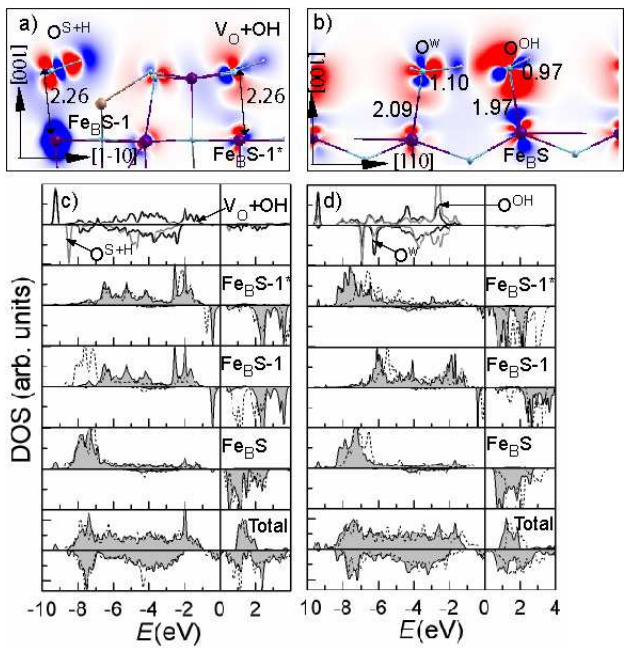}}
  \caption{ Adsorbate induced electron density redistribution for (a) B+V$_{\rm O}$+1\hto(D) and (b) B+4\hto(M). Red/blue  corresponds to regions of charge accummulation/depletion.
  Relevant bond lengths are given in {\AA}. Total and projected density of
  states (DOS) before (dashed line) and after water adsorption (solid line,
  grey area) for (c) B+V$_{\rm O}$+1\hto(D) and (d) B+4\hto(M).  S (S-1) denote surface (subsurface) \feb\ sites.}
  \label{dos}
\end{figure}

We have compared the stability of more than 30 different configurations
using \abt~\cite{reuter}. The surface energy,
$\gamma(T,p) = \frac{1}{2A}\left( G_{\rm Fe_3O_4(001)}^{slab}
-N_{\rm Fe}\mu_{\rm Fe}-N_{\rm O}\mu_{\rm O} -N_{\rm H_2O}\mu_{\rm H_2O} \right) $
depends on the Gibbs free energy and the chemical potentials of the
constituents. The Gibbs free energy can be expressed
through the total energy from the DFT-calculations~\cite{reuter}.
Because there are two species in the gas phase, O$_{2}$ and \hto,
the surface phase diagram is three dimensional.
Fig.~\ref{surf_phase} displays a two dimensional projection with the
most stable configurations for given $\mu_{\rm O}$, $\mu_{\rm
H_2O}$. We first consider the termination of the clean surface as a
function of oxygen pressure. As discussed previously, a modified
B-layer (denoted as B) showing lateral and vertical
distortions in the surface layer with a {\sl wave-like} pattern is
favored over a broad range of oxygen
pressures. However, at oxygen poor conditions a B-layer with oxygen
vacancies (B+V$_{\rm O}$) is stabilized. This defective surface, previously
proposed in a STM study~\cite{stanka00}, shows dramatic relaxations
where the oxygen opposite the vacancy moves towards the \feb-row by
$\sim$0.8 \AA. Starting from these two
terminations~\cite{Alayer}, we have adsorbed water molecules, varying
their concentration and geometry. We find a strong tendency for isolated molecules to
dissociate in surface oxygen vacancies (V$_{\rm O}$+OH) whereby a
proton diffuses to a surface oxygen (\Os) further away and forms a
surface OH-group (O$^{\rm S}$+H). Thus, even at very low water
vapor pressures, all surface defects are expected to be filled with
OH-groups, consistent with XPS results~\cite{Kendelewicz}. The
dissociation of water in the defect sites invokes significant
changes in the electronic structure as shown in Fig.~\ref{dos}a and
c, involving the $1\pi$ and $3\sigma$ molecular orbitals of the OH
groups and a switching from \feth\ to \fetw\ of \feb S-1 underneath
O$^{\rm S}$+H.

The adsorbate-adsorbate interaction that sets in, when a second
molecule is adsorbed within the \sqtwo~unit cell, leads  to a mixed
adsorption mode: One molecule dissociates protonating a surface
oxygen, while the intact molecule forms a hydrogen bond with the
OH-group.
The main part of the phase diagram is dominated by a mixed
adsorption mode of four water molecules where all surface \feb-sites
(\feb S) are saturated (B+4$\rm H_2O(M)$). Full dissociation  is
12~meV/\AA$^2$ less stable. Full hydroxylation of the surface
is also not likely to occur as  the
formation of a surface OH-group neighboring a subsurface \fea\ is
extremely unfavorable.

The electron density redistribution with respect to the B-layer in Fig.~\ref{dos}b
indicates a weak charge accumulation between \feb S and the water
molecule (\Ow), but the strongest charge rearrangement takes
place between \feb\ and the OH-group (\Ooh) of the dissociated
molecule, involving depletion of $3\sigma$ and accumulation in
$1\pi$ molecular orbitals of OH along with depletion of the $d_{z^2}$
orbital at \feb S.
The main actuator of partial dissociation appears to be the
formation of strong intermolecular hydrogen bonds.
As a result both \hto\ and OH tilt from the on-top position
towards each other and \Ow-H elongates from 0.95~\AA\ (gas phase) to 1.10~\AA,
resulting in an \Ow-H$\cdot\cdot\cdot$\Ooh\ of 2.47~\AA. A
fingerprint of the mixed adsorption are the two distinct bond
lengths between \feb-\Ow\ and \feb-\Ooh\ of 2.09~\AA\ and 1.97~\AA\,
respectively. This feature is confirmed by the LEED  analysis described below.

\begin{figure}[bt]
  \centering
  \scalebox{0.35}{\includegraphics{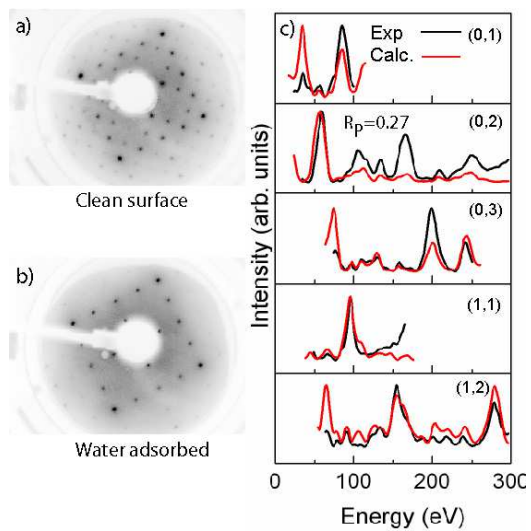}}
  \caption{(Color online) LEED pattern a) before and b) after water adsorption. The superstructure spots in a)
  indicative of the \sqtwo surface reconstruction are largely suppressed in b).
  c) experimental (averaged over symmetrically equivalent beams - black) and calculated (red/grey) LEED $I(V)$ curves of the water adsorbed surface.}
  \label{fig:leedpat}
\end{figure}

LEED measurements were performed on a synthetic magnetite
sample. The clean surface, prepared by Ar$^+$-ion sputtering and
subsequent annealing at 900-1000 K at $p_{\rm O_2}=5\times 10^{-7}$
mbar for 2-3 h, exhibited a \sqtwo-LEED pattern  with sharp
superstructure reflections and low background
(Fig.~\ref{fig:leedpat}~a).  Water was adsorbed at 273~K for 2 min
at a water vapor pressure of $2\times
10^{-6}$ mbar, resulting in a $(1\times
1)$-LEED pattern with an enhanced background
(Fig.~\ref{fig:leedpat}b). While annealing up to 770~K restores the superstructure spots, the shape of the the LEED $I(V)$ curves cannot be recovered~\cite{leed}, indicating some residual hydroxylation of the surface, as observed by Kendelewicz \etal~\cite{Kendelewicz}.
Seven LEED $I(V)$ curves of the $(1\times 1)$-surface  were measured in the energy range 50 -
300 eV at 100~K. An increase of the background current during the measurement is attributed to uncorrelated defects induced by the electron beam~\cite{beam}. However, no changes were observed in the LEED $I(V)$ curves  during repeated
measurements over periods of several hours.

LEED calculations were performed with  the layer doubling method and
a least squares optimization~\cite{Over92} using constraints for bond lengths.
The crystal potential was calculated from a
superposition of atomic potentials using optimized muffin-tin radii
~\cite{rundgren_03} which led to reliable structural determination of the clean
\magn(001)-surface~\cite{pentcheva08}. 10 phase shifts were used. All
positions and occupation numbers within the adsorbate and top B-A-B substrate layers
were optimized in a \sqtwo-unit cell, while thermal parameters were kept fixed.
The best fit ($R_P$ = 0.27) was obtained for a model where all \feb S
sites have adsorbed oxygen on top. Both the surface \feb- and
adsorbate sites are occupied by $\sim 80\%$ possibly due to defect creation
during the preparation procedure.
The  adsorbed O shows strong lateral shifts  by $\sim 0.4$~\AA\ off the \feb\ sites in
agreement with the DFT results (0.23-0.28~\AA).  The
main feature are the two different bond lengths \feb-\Ow/\Ooh:
2.12~\AA\ and 1.93~\AA, confirming the simultaneous occurrence of
hydroxyl groups and molecular adsorption. Further details on the
structural analysis will be published elsewhere~\cite{leed}.

\begin{figure}[bt]
 \centering
 \scalebox{0.45}{\includegraphics{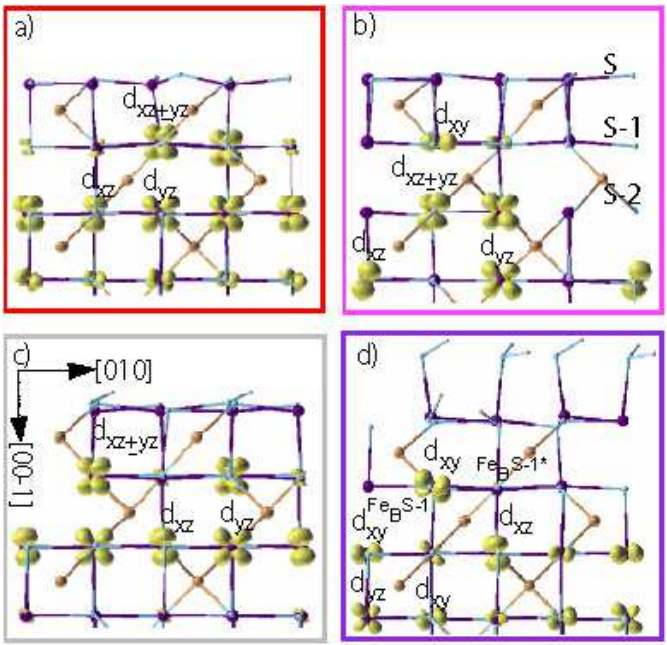}}
 \caption{Side view of the electron density integrated between -1.3 eV and $E_{\rm F}$ showing the occupation of the minority $t_{2g}$ orbitals at the \feb-sites. a) B+V$_{\rm O}$; b) B-layer; c) B+V$_{\rm O}$+1\hto (D); d) B+4\hto(M). For the colour code see Fig.~\ref{surf_phase}.}
 \label{fig:co-oo}
\end{figure}

We turn next to the surface and adsorbate induced electronic effects
on \magn(001). The electron density plot in Fig.~\ref{fig:co-oo}
displays the minority $t_{2g}$ occupancies  at the
\feb-sites, thus allowing to distinguish between sites with
predominant \fetw~or \feth-character~\cite{Uvalue}. The notation \fetw\ and
\feth\ is used here for simplicity, but the difference in total
$3d$ occupation within the muffin-tin sphere  is much smaller
(0.2-0.4$e$) consistent with
XRD-~\cite{radaelli} and LDA+$U$ studies for the low temperature
bulk phase~\cite{leonov,guo}. The magnetic moments allow a
clearer discrimination: $M_{\rm Fe^{2+}}=3.54-3.75 \mu_{\rm B}$,
$M_{\rm Fe^{3+}}=3.90-4.10 \mu_{\rm B}$, respectively. For most of the
stable systems we find that the surface layer contains exclusively
\feth\, while a unique charge and orbitally ordered state emerges in
the deeper layers depending on the type of termination~\cite{Ovac}.
In B+V$_{\rm O}$ (Fig.~\ref{fig:co-oo}a) the two subsurface \feb\ next to V$_{\rm O}$  are \fetw\ ($d_{xz}\pm d_{yz}$).
Upon water adsorption the positions of \fetw\
switch in B+V$_{\rm O}$+1\hto(D): now the two \feb\ below
the surface OH groups are \fetw, resulting in alternating \fetw,
\feth-sites. In the mixed adsorption case (B+4\hto(M))
the charge ordering in the subsurface layer is not significantly
altered compared to B-layer, where  one out of
four \feb S-1 is  \fetw\
and a second is in an intermediate valence state.
However, the slight tilting of the $t_{2g}$ orbital at \feb $^{2+}$  induces a
completely different OO  in S-2: from $d_{xz}\pm d_{yz}$ to
alternating $d_{xy}$ and $d_{xz}$-orbitals. Recent GGA+$U$
calculations have shown that in bulk magnetite a variety of CO/OO
states can be realized  by symmetry lowering~\cite{pinto06}. Here
the presence of a surface and adsorbates imposes a unique CO/OO
state for each termination reaching several layers below the
surface. Moreover, most of them violate the Anderson criterion:
Fe$_4$O$_4$ cubes in the surface and subsurface layer are
predominantly electron poor (3\feth+1\fetw), while those in S-1 and
S-2-layer are electron rich (1\feth+3\fetw). Evidence for orbital
ordering in thin magnetite films was reported recently from  resonant
soft x-ray diffraction~\cite{schlappa}.

The charge ordered state in the B-layer leads to the opening of a
band gap of 0.3~eV as shown in Fig.~\ref{dos}d, consistent with
previous calculations~\cite{lodziana07} and scanning tunneling spectroscopy
measurements~\cite{jordan06}. However the adsorption of water on the
surface and the formation of surface hydroxyl groups leads to half-metallic behavior.

We have shown that on \magn(001) water tends to
dissociate in oxygen vacancies or partially dissociate on the
nondefective surface in contrast to the full dissociation found on
Fe$_3$O$_4$(111)~\cite{Kendelewicz,magnetite_111}. This adsorption
mode is triggered by the presence of both Lewis acid and base sites on the surface and the intermolecular interaction as reported also
on rutile(110)~\cite{TiO2_PRL}, anatase(001)~\cite{TiO2_ana} and MgO(001)-surfaces~\cite{MgO_PRL}. The hydrogen bonded OH and \hto\ can easily
rearrange and thus provide adsorption sites for further species
(\eg\ heavy metal complexes). Our results indicate a  pathway to manipulate
\eg\ the  catalytic properties of transition metal oxide surfaces by triggering
electron transfer processes  and inducing new charge and orbitally ordered states via adsorbates.

We acknowledge discussions with G.E. Brown and T. Kendelewicz and support by the German Science Foundation (PE883/1-2) and a grant for computational time at the Leibniz Rechenzentrum Garching.


\begin{thebibliography}{10}
\bibitem{Uvalue} We have varied $U$
between 0-8 eV: Charge disproportionation at the
\feb-sites is found beyond $U=2$~eV.
\bibitem{calcdet}We have used muffin tin (MT)
radii of $R^{MT}_{H}=0.60$ Bohr, $R^{MT}_{Fe}=1.90$, and
$R^{MT}_{O}=1.10$ and an energy cutoff for the wave functions and
potential of $E^{wf}_{max}=25$~Ry and $E^{pot}_{max}=196$~Ry,
respectively.  For the integration in the Brillouin zone 16
$k{\parallel}$-points  were used.
\bibitem{Alayer} A clean or water adsorbed
half-occupied A-layer was found to be less stable in the pressure
ranges studied here.
\bibitem{beam} For a filament current of 50 -100 nA a rough estimate
of the electron dose  is 1-2 (e/u. c.) per 10 s. 
\end{thebibliography}
\end{document}